# Quantum spin Hall insulators and quantum valley Hall insulators of BiX/SbX (X = H, F, Cl, and Br) monolayers with a record bulk band gap

Zhigang Song[1†], Cheng-Cheng Liu[2†], Jinbo Yang[1,3*], Jingzhi Han[1], Meng Ye[1], Botao Fu[2], Yingchang Yang[1], Qian Niu[3,4], Jing Lu[1,3 *] & Yugui Yao[2 *]

[1]State Key Laboratory for Mesoscopic Physics and School of Physics, Peking University, Beijing 100871, China

[2]School of Physics, Beijing Institute of Technology, Beijing 100081, China

[3]Collaborative Innovation Center of Quantum Matter, Beijing, China

[4]International Center for Quantum Materials, Peking University, Beijing 10087, China

Keywords: topological insulator; quantum valley Hall insulator; functionalized Bi and Sb monolayers; large band gap

Abstract

Large bulk band gap is critical for application of the quantum spin Hall (QSH) insulator or two dimensional (2D) topological insulator (TI) in spintronic device operating at room temperature (RT). Based on the first-principles calculations, here we predict a group of 2D topological insulators BiX/SbX (X = H, F, Cl, and Br) monolayers with extraordinarily large bulk gaps from 0.32 to a record value of 1.08 eV. These giant-gaps are entirely due to the result of strong spin-orbit interaction related to $p_x$ and $p_y$ orbitals of Bi/Sb atoms around the two valley $K$ and $K'$ of honeycomb lattice, which is different significantly from the one consisted of $p_z$ orbital just like in graphene/silicene. The topological characteristic of BiX/SbX monolayers is confirmed by the calculated nontrivial $Z_2$ index and an explicit construction of the low energy effective Hamiltonian in these systems. We show that the honeycomb structures of BiX monolayers remain stable even at a temperature of 600 K. These features make the giant-gap TIs BiX/SbX monolayers an ideal platform to realize many exotic phenomena and fabricate new quantum devices operating at RT. Furthermore, biased BiX/SbX monolayers become a quantum valley Hall insulator, showing valley-selective circular dichroism.

**Introduction**

The quantum spin Hall (QSH) insulators, also known as two-dimensional (2D) topological insulators (TIs), have generated great interest in the fields of the condensed matter physics and materials science due to their scientific importance as a novel quantum state and potential applications in ranging from spintronics to topological quantum computation [1-3]. The QSH insulators are characterized by an insulating bulk and fully spin-polarized gapless helical edge states without backscattering at the sample boundaries, which are protected by time-reversal symmetry. The prototypical concept of the QSH effect was first proposed by Kane and Mele in graphene in which the spin-orbit coupling (SOC) opens a band gap at the Dirac point [4,5]. However, the rather weak second order effective SOC makes the QSH effect in graphene only appear at an unrealistically low temperature [6].

Up to now only the HgTe/CdTe quantum well is verified to be a well-established QSH insulator experimentally [7,8]. Experimental evidence has also been presented recently for helical edge modes in inverted InAs/GaSb quantum wells [9]. The critical drawback of such reported QSH state is their small bulk gaps, which are too small to make the predicted QSH effect observable under experimentally easy accessible conditions. Thus, to observe QSH effect at room temperature (RT) in TIs, large bulk band gap is essential because they can stabilize the edge current against the interference of the thermally activated carriers in the bulk due to the fact that the carrier concentration in the bulk decreases exponentially with the band gap. Extensive effort has been devoted to search for new 2D TIs with a large bulk band gap [10-14]. Some layered materials such as silicene, germanene [15] and stanene [16] have been proposed, and the bulk band gap of 2D TI has been elevated to remarkable 0.3 eV in chemical modified tin film, SnX (X = F, Cl, Br, and I) [13]. Recently, ultrathin Bi films have drawn much attention as a promising candidate of the QSH insulator, and the 2D topological properties of the ultra-thin Bi(111) films have been reported [17]. To the best of our knowledge, no bulk band gap has exceeded 0.7 eV in both 2D and 3D TIs [18].

Since Bi and Sb are well known for their strong SOC that can drive and stabilize the topological non-trivial electronic states, it is wise to search for large-band-gap QSH insulators based on the Bi/Sb related materials. Here, we predicted that the free-standing 2D honeycomb

Bi/Sb halide and Bi/Sb hydride (We call them bismuthumane and stibiumane, respectively, by analogy with graphane, silicane, and stanane) systems are stable huge-band-gap QSH insulator based on the first-principles (FP) calculations of the structure optimization, phonon modes, and the finite temperature molecular dynamics as well as the electronic structures. The topological characteristic of these TIs is confirmed by the FP-calculated nontrivial $Z_2$ index. The low-energy effective Hamiltonian (LEEH) is given to capture the low-energy long-wavelength properties of these systems. Significantly, among these new TIs, we found that the bulk band gap of about 1.0 eV related to the $p_x$ and $p_y$ orbitals of the Bi atoms in BiX (X = H, F, and Cl) monolayers. To our knowledge, these are the largest-band-gap TIs. Their gaps opened by SOC in QSH phase can be effectively tuned by the X atom. All of the above features make these compounds promising for the applications at RT. Moreover, when the inversion symmetry of the honeycomb lattice for BiX/SbX monolayers is broken, BiX/SbX monolayers become a quantum valley Hall insulator, and chiral optical selectivity of the valleys is obtained.

**Methods**

For these materials, we first carried out a geometry optimization including SOC interaction using the VASP package within the framework of the projector augmented wave (PAW) pseudopotential method using a plane-wave basis set. The Brillouin-zone integrations have been carried out on a $9\times9\times1$ Γ-centered $k$ mesh. Vacuum regions with thickness larger than 14 Å were placed to avoid interaction between the monolayers and its periodic images. Both the atomic positions and lattice constant were relaxed until the maximal force on each relaxed atom was smaller than 0.001 eV/Å. The cutoff energy for wave-function expansion was set as 1.3* $E_{max}$ of the X atoms. The stability of the optimized structure for BiH monolayer was confirmed by a vibrational analysis using the phonopy package [19] with a supercell of $5\times5$ unit cells. Fully relativistic band calculations were performed with the LAPW (linearized augmented plane wave) method implemented in the WIEN2K package, and the results are in good agreement with those generated by the VASP package. SOC was included as a second vibrational step using scalar-relativistic eigenfunctions as basis after the initial calculation being converged to self-consistency. The relativistic $p_{1/2}$ corrections were also considered for 6$p$ orbital of Bi in order to improve the accuracy. A $20\times20\times3$ $k$-points grid was utilized in the first Brillouin zone sampling and cutoff parameters $R_{mt}\cdot K_{max}$ were 4 for BiH/SbH monolayers and 6 for BiX/SbX (X = F, Cl, and Br) monolayers, respectively. The Fermi energy was calculated where each eigenvalue was temperature broadened using a Fermi function with a broadening parameter of 0.002 Ry. The exchange-correlation functional was treated using Perdew-Burke-Ernzerhof generalized gradient approximation throughout the paper.

**Results**

Figure 1(a) plots the typical optimized geometries for BiX monolayers, which have a three-fold rotation symmetry like that in graphene. The inversion symmetry holds for all tested compounds. The obtained equilibrium lattice constants, nearest neighbor Bi-X distances and buckling heights through structural optimization were listed in Table 1. A quasi-planar geometry is found to be more stable for BiH monolayer (bismuthumane), while a low-buckled configuration is more stable for BiF, BiCl, and BiBr monolayers. This is related to the bonding between Bi and the X atoms. Since F, Cl, and Br are more electronegative than H, the bond between Bi and F atoms is stronger than that between Bi and H atoms, leading to a low buckling in BiX (X = F, Cl and Br) monolayers. The lattice constants of BiX monolayers follows the sequence of $a$(F) < $a$(Cl) < $a$(Br), in accordance with the electronegativity. The bond distances of the Bi-X films slightly increases with the sequence of $d$(Br) > $d$(Cl) > $d$(F) determined by their covalent bond radii. The kinetic stability of these 2D TIs is further confirmed by the calculations of the phonon spectrum without SOC. Taking BiH monolayer for example (Fig. 1(c)), there is no imaginary frequency along all momenta, which indicates that this structure is kinetically stable, corresponding to an energy minimum in the potential energy surface.

Thermodynamical stability of BiX/SbX monolayers is then checked by calculating the per-atom Gibbs free energy of formation ($\delta G$),

$$\delta G = E + n_{Bi}\mu_{Bi} + n_X\mu_X \tag{1}$$

where $-E$ represents the cohesive energy per atom of the BiX/SbX monolayers, $n_{Bi}$ and $n_X$ are the mole fractions of Bi and X atoms, respectively, for a given structure, and $\mu_{Bi}$ and $\mu_X$ are the per-atom chemical potentials of Bi and X, respectively, at a given state. We chose $\mu_{Bi}$ and $\mu_X$ as the binding energies per atom of bulk Bi, and $X_2$ molecule, respectively. We provide the formation energy data of all the checked BiX/SbX monolayers in Table 2. The calculated $\delta G$ value for BiH monolayer is 0.30 eV. Bismuthine is a chemical compound with the formula $BiH_3$. It is stable below −60 ℃ [20,21]. $\delta G$ of bismuthine is 0.53 eV. $\delta G$ of BiH monolayer is smaller than $\delta G$ of Bismuthine, therefore it is possible to synthesize BiH monolayer. Remarkably, bismuth/antimony-halide monolayers have negative $\delta G$, indicating a higher thermodynamical stability relative to their elemental reservoirs.

We carried out *ab initio* molecular dynamics (MD) simulations using a supercell of $3\times3$ unit cells at various temperatures (see Fig. 2 and Fig. S1) with a time step of 1.5 fs to check thermal stability of BiX monolayers. After running 1500 steps at 300 and 600 K, no bond is broken, suggesting that the structures of BiX (X = H, F, Cl, and Br) monolayers are thermally stable even at a temperature of 600 K. We also performed an *ab initio* MD simulation for a larger $4\times4$ supercell for BiH monolayer and found that the structure of BiH monolayer is stable after 2.25 ps at 600 K (See Fig. S2). In fact, it was found that the Bi-X bond energy is much higher than that of Bi-Bi bonds due to a large bond distance between Bi-Bi atoms. The snapshots of the MD simulations at higher temperature show that the Bi-Bi bonds are broken while Bi-X bonds remain at 700 K. SbX monolayers are also stable at 300-400 K (see Fig. S1). The thermal stability of these structures enables these films to be used at or even above RT, which is very important for the practical applications.

The typical band structures of the predicted systems BiH, BiF, and SbF are shown in Fig. 3(a-c). The band structures of other monolayers are provided in Fig. S3. The valence and conduction bands near Fermi level are mostly composed of the $p_x$ and $p_y$ orbitals from the Bi atoms according to the partial band projections. Notably, the two energy bands are shown to cross linearly at the $K$ (and $K = -K'$) point, suggesting the existence of Dirac-cone-like features in the band structure of these two-dimensional honeycomb systems without SOC. It means that these materials can be considered as a gapless semiconductor, or alternatively, as a semi-metal with zero density of states (DOS) at Fermi level. Because the honeycomb structure consists of two equivalent hexagonal Bi sublattices, the electrons in these predicted materials can formally be described by a Dirac-like Hamiltonian operator containing a two-component pseudospin operator.

When SOC is switched on, the degeneracy at the Dirac points is lifted, and the valence bands are down shifted while the conduction bands are up shifted, producing a huge band gap opened by SOC for all BiX monolayers. The local gap at the Dirac point $K$ ($K'$) is a result of the first order relativistic effect related to Bi elements. Thus the gap is robust. On the contrary, the conduction bands are down shifted while the valence bands are up shifted at the $\Gamma$ point, which produces a global indirect band gap. The X atoms mainly hybridize with Bi atom near the $\Gamma$ point in the conduction and valance bands. The band gap can be effectively tuned by the X atoms. The global gaps of BiX (X = H, F, Cl and Br) monolayers are in the range from 0.74 to 1.08 eV owing to the strong SOC of the Bi atoms, especially for BiH and BiF monolayers with bulk gaps larger than 1.0 eV. As for SbX (X = H, F, Cl, and Br) monolayers, the valence bands are down shifted while the conduction bands are up shifted at the $K$ point, producing a global direct band gap. The values of the gaps are in the range from 0.3 to 0.4 eV, which are comparable to those of the theoretically predicted chemically modified tin films [13].

The band topology of BiX/SbX (X = F , Cl, and Br) monolayers can be characterized by the $Z_2$ invariant [22]. $Z_2 = 1$ characterizes a nontrivial band topology (corresponding to a QSH insulator) while $Z_2 = 0$ means a trivial band topology. The $Z_2$ invariants can be directly obtained from the FP lattice computation method [23]. Taking BiH monolayer for example, the $n$-field configuration is shown in Fig. 3(d) from FP calculations. It should be noted that different gauge choices result in different $n$-field configurations; however, the sum of the $n$ field over half of the Brillouin zone is gauge invariant module 2, namely $Z_2$ topological invariant [24]. The honeycomb BiH monolayer has nontrivial band topology with the topological invariant $Z_2 = 1$, and at the Dirac point $K$ the gap opened by SOC is sizable. We also calculate $Z_2$ topological invariant for the other systems, and find that they are all topological non-trivial. Therefore, the QSH effect can be steadily realized in the 2D honeycomb Bi(Sb) hydride/halide with huge SOC gap.

To the best of our knowledge, a bulk band gap of over 1.0 eV in BiH and BiF monolayers is the largest bulk band gap of all the reported TIs. The band gaps of these compounds are about 3 times of the recent results of theoretically predicted chemical modified tin films (a bulk gap of 0.3 eV) [13] and the superstar 3D topological insulator $Bi_2Se_3$ (a bulk gap of 0.35 eV) [25]. Furthermore, the predicted large bulk gap makes BiX/SbX monolayers capable of enduring considerable crystal defects and thermal fluctuation which are beneficial to the applications in high-temperature spintronics device. Bi is among the main group elements that have the strongest SOC, a fundamental mechanism to induce the $Z_2$ topology. For this reason, the predicted TIs consisted of Bi show huge gap opened by SOC.

Bulk band gap is one of the most important parameters for TIs and is in analogy to superconducting transition temperature ($T_c$) for superconductor. Insulation of the bulk is critical to observe the surface metallic state of a TI because the surface metallic state would be masked if the bulk state becomes metallic. A large bulk band gap is critical to maintain bulk insulating. If the bulk band gap is too small, the defect and disorder, which are difficult to avoid in material growth, would probably shift the Fermi level to the conduction or valence band, making the bulk conductive. Besides, if the bulk band gap is too small, carrier (electron and hole) can be more easily produced at a finite temperature. The generation of the TIs can be categorized chiefly in terms of their bulk band gap values [1,26]: The first generation TIs includes Bi-Sb alloy with a bulk gap smaller than 0.1 eV, and the second generation includes $Bi_2Se_3$, $Sb_2Te_3$, and $Bi_2Se_3$. The previously reported TIs SnX (X = F, Cl, Br, and I) can also be categorized into this generation. Our reported TIs BiX (X = H, F, Cl, and Br) monolayers have a bulk band gap of about 1.0 eV and can be regarded as the third generation TIs. Therefore, although the idea to realize a new TI by functionalization of a 2D material in this paper is similar to that in the previous work, our results predict a new generation of TIs and stand for an important breakthrough in TIs study.

Large lattice distortion really affects the energy band and the band gap of BiH monolayer. Based on the MD calculation, we predicted that phase transition temperature of BiH monolayer is between 600 and 700 K. At a temperature of 600 K, the lattice distortion is very large, and the inversion symmetry is destroyed (See Fig 2). However, the band gap remains larger than 0.22 eV after 2.25 ps. The bands are split by SOC in absence of inversion symmetry, and the splitting is mainly a Rashba type. We calculated the $Z_2$ number of the structure of the MD simulation at 600 K after 2.25 ps and found that BiH monolayer remains a TI. Hence, the topology of BiH monolayer is very robust against a lattice distortion.

We also studied the effects of inversion symmetry breaking induced by an electric field on the band gap and topology of BiH monolayer. It was found that the band gap decreases with the increasing electric field (See Fig S4) and closes at $E$ = 0.67 V/Å, but the nontrivial topology remains. It is noteworthy that under a certain range of electric fields (0.61-0.67 V/Å), BiX monolayers may generate perfect free electron gas and serve as a spin field effect transistor (See details in Supplementary Information ( I )).

The nontrivial topology of SnX monolayers in the previous work origins from *s*- and *p*-band inversion at the Γ point similar to that in HgTe quantum well and in $Bi_2Se_3$ [8,25]. However, the nontrivial topology in BiX/SbX monolayers results from the massive Dirac cone, and there is no band inversion. Actually, the origin of nontrivial topology in BiX/SbX monolayers is similar to that in graphene and silicene, but the type of SOC in BiX/SbX monolayers is brand new. We construct a minimal model Hamiltonian on the basis of FP calculations and general symmetry consideration. The symmetry of these systems possesses $D_{3d}$ point group and the groups of the wave vector at the Dirac points $K$ and $K'$ are both $D_3$, which splits the *p* orbitals at the Dirac points into two groups: $A_2$ ($p_z$) and $E$ ($p_x$, $p_y$). Based on the FP calculations, around the Dirac points and Fermi level, the low-energy band structure is mainly consisted of $p_x$ and $p_y$ orbitals from Bi/Sb atoms in the band components. The $p_x$ and $p_y$ orbitals make up 2D irreducible representation of the wave vector of $D_3$ at the Dirac points, which is relevant for the low-energy physics. There are massive Dirac cones at the $K(K')$ point and flat bands (the second band below the Fermi level) [27] in BiX/SbX monolayers. Massive Dirac cones and flat bands mainly consisting of the $p_x$ and $p_y$ orbitals distinguish BiX/SbX monolayers from graphene/silicene and lead to new phenomena, such as orbital analogue of the quantum anomalous Hall effect [28] and Wigner crystallization [36].

Taking into account that there are A and B two distinct sites in the unit cell (Fig. 1(a)), the symmetry-adopted basis functions can be written as $|\Phi_1\rangle = -\frac{1}{\sqrt{2}}\left(p_x^A + i\tau_z p_y^A\right)$, $|\Phi_2\rangle = \frac{1}{\sqrt{2}}\left(p_x^B - i\tau_z p_y^B\right)$, with $\tau_z$ labeling the valley degree of freedom, K and K', which means that the basis functions are different around the K and K' points. SOC term generally reads $H_{SO} = \xi_0 \vec{L}\cdot\vec{S} = \xi_0\left(\frac{L_+ S_- + L_- S_+}{2} + L_z S_z\right)$, where $S_\pm = S_x \pm i S_y$ and $L_\pm = L_x \pm i L_y$ denote the creation (annihilation) operator for the spin and angular momentum, respectively. $\xi_0$ is the magnitude of effective intrinsic SOC. A straightforward calculation leads to the on-site SOC in the spinful low-energy Hilbert subspace $\{|\Phi_1^\uparrow\rangle, |\Phi_1^\downarrow\rangle, |\Phi_2^\uparrow\rangle, |\Phi_2^\downarrow\rangle\}$, $H_{SO} = S_z \lambda_{SO} \tau_z \sigma_z$, with $\lambda_{SO} = \frac{1}{2}\xi_0$. The low-energy Hilbert subspace consisting of $p_x$ and $p_y$ orbitals differs significantly from the one consisting of $p_z$ orbital just like in graphene and silicene. Moreover, the SOC term is on-site rather than the next nearest neighbor as in Kane-Mele model [4,5,15,16]. This indicates that the SOC mechanism in BiX/SbX monolayers is totally different from that in the graphene or silicene.

To the first order of k, the symmetry-allowed four-bands LEEH involving SOC can be written as,

$$H = \hbar v_F (k_x \sigma_x + \tau k_y \sigma_y) + S_z \lambda_{SO} \tau_z \sigma_z, \qquad (2)$$

where Pauli matrix $\sigma$ denotes $|\Phi_1\rangle$ and $|\Phi_2\rangle$ orbital degree of freedoms, and $\tau_z$ labels the valley degree of freedom K and K'. The energy spectrum of the total LEEH is $E(\vec{k}) = \pm\sqrt{\hbar^2 v_F^2 k^2 + \lambda_{SO}^2}$ with a gap $E_g = 2\lambda_{SO}$ at the Dirac Points. The above LEEH is invariable under the space reversal and the time reversal operation. It should be noticed that in fact the low-energy basis functions are mixed with small components of other orbitals (See Fig. 3), whereas the low-energy physics can be grasped by the low-energy effective model. The only two parameters $v_F$ and $\lambda_{SO}$ in the above effective Hamiltonian can be obtained from FP calculations, whose values are listed in Table 1. The band structures around the *K* point for BiH, BiF, and SbF monolayers calculated using DFT and LEEH methods are compared in Fig. 4. It is obvious that in the vicinity of the Dirac *K* point there is a good agreement between the calculated results of these two methods.

There is a valley degree of freedom in BiX/SbX monolayers, and the SOC opens a large gap at corners of the Brillouin zone. If the inversion symmetry is broken in BiX/SbX monolayers, for example by a vertical electric field, the valley-contrast Berry curvature appears and they become a quantum valley Hall insulator. By contrast, SnX monolayers are not a quantum valley Hall insulator at all (the band gap appears at the Γ point) [13]. The valley-contrast Berry curvature and spin in BiX/SbX monolayers will give rise to novel transport properties, such as valley Hall effect [29], valley spin Hall effect [30] and valley orbital moment Hall effect. Besides, valley-contrast circular dichroism will appear. Generally, degree of valley polarization $\eta(k)$ and circular dichroism can be described by degree of circular polarization $\eta(k)$ in Brillouin zone. $\eta(k)$ is defined as $\eta(\boldsymbol{k}) = \frac{|p_+(\boldsymbol{k})|^2 - |p_-(\boldsymbol{k})|^2}{|p_+(\boldsymbol{k})|^2 + |p_+(\boldsymbol{k})|^2}$, where $P_\pm = P_x \pm iP_y$; $p_\alpha$ is the matrix element between the conduction and valence bands and is given by $P_\alpha = \langle u_v(k)| \frac{1}{\hbar}\frac{\partial H}{\partial k_\partial} |u_c(k)\rangle$. $|u_c(\boldsymbol{k})\rangle$ and $|u_v(\boldsymbol{k})\rangle$ are the periodic parts of the conduction and valence band of Bloch function, respectively. The calculated $\eta(k)$ in irreducible Brillouin zone of BiH monolayer under a vertical field is shown in Fig. 5, and a perfect optical selection rule at two valleys is apparent. Namely, the valley *K* absorbs left-handed photons, while the valley *K'* absorbs right-handed photons. Thus a circular polarized light can be used to create imbalanced electron occupation between the two valleys in BiX/SbX monolayers. However, in graphene and silicene, valleytronics is difficult to realize due to the quite small SOC.

Moreover, in BiX/SbX monolayers, there is a new and strong coupling between spin and valley pseudo-spin due to the large band gap opened by SOC, which is different from that in the MoS$_2$ system. The valley pseudo-spin can be controlled using the electric field due to the strong spin-orbital coupling in the BiX/SbX monolayers. According to the low power model, there is intrinsic valley degree of freedom and the valley orbital moment couples to the spin. The low energy effective model with broken inversion symmetry in BiX/SbX monolayers is:

$$H = \hbar v_F(k_x\sigma_x + \tau k_y\sigma_y) + S_z\lambda_{SO}\tau_z\sigma_z + \frac{\Delta_1}{2}\sigma_z, \qquad (3)$$

where $\Delta_1$ is the additional band gap induced by inversion symmetry breaking. In the low energy limit, the valley magnetic moments of BiX/SbX systems can be expressed as: $m = \tau\frac{e\hbar v_F^2}{2s_z\tau_z\lambda_{SO} + \Delta_1}$, where $s_z$ is the real spin. By contrast, in the MoS$_2$ monolayer, the valley moments are: $m = \tau_z\frac{ev_F^2}{2\hbar\Delta_1}$ [31]. It can be seen that the valley magnetic moments of BiX/SbX monolayers are related to not only the valley pseudo-spin but also the real spin. $\Delta_1$ is a variable which can be adjusted by the electric field.

**Discussion**

Chemical functionalization of 2D materials is a powerful tool to create new materials with desirable features, such as graphane or fluorinated graphene. In the current study, we have investigated the properties of the bismuth monolayer with planar or low-buckled structure, and found its structure unstable without X atom. The high-buckled Bi monolayer, i.e., bilayer bismuth film with a lattice constant of 4.52 Å is more stable than the bismuth planar monolayer with a lattice constant of about 5.4 Å [11]. The high-buckled bismuth monolayer compounded with X elements may increase their crystal lattice constants by about 1.0 Å, resulting in a quasi-planar or low-buckled monolayer configuration. BiX monolayers show totally different band structures from bismuth monolayer characterized by a doubled bulk band gap. Actually, we found that the chemical functionalization of As, P, and N monolayers can also result in 2D TIs with bulk band gaps of 0.18, 0.03, and 0.01 eV, respectively. Therefore functionalization is an effective approach to obtain 2D TIs. On the experimental side, it has been known that stable halides of Bi and Sb such as BiX/SbX (X = F, Cl, Br, and I) [32,33] have been synthesized although hydrides of Bi and Sb are unstable at RT [20,21].

Four different preparation methods were proposed here: (**a**) The method to synthesize graphane [34] may be applied to prepare BiH monolayer. High-buckled monolayer bismuth can be prepared by peeling off the bulk bismuth [17]. Bismuth monolayer should be first annealed in an argon atmosphere in order to remove any possible contamination and then exposed to the cold hydrogen plasma [34]. (**b**) Before hydrogenation, monolayer bismuth samples are heated in vacuum to remove physisorbed polymers and other contaminants that might block the hydrogenation of the graphene surface. Monolayer bismuth can be exposed to hydrogen gas in a suitable condition. Hydrogenation can be performed in a vacuum system equipped with a hot tungsten filament that can split the $H_2$ gas into hydrogen atoms [35,36]. (**c**) Monolayer Bismuth samples are deposited on Si wafer. Then, a layer of hydrogen silsesquioxane is coated on the bismuth samples and irradiated electrons at various doses [37]. Hydrogen atoms are generated in situ by breaking Si-H bonds of hydrogen silsesquioxane in the course of e-beam lithography. Finally, BiH monolayer is formed. (**d**) According to the exfoliation mechanism of monolayer BN using molten hydroxides [38], we might expose bulk bismuth to a cold hydrogen plasma to grow BiH monolayer. When Bi atoms on the surface of bulk bismuth are combined with hydrogen atoms, the Bi-Bi bonding becomes longer, and bismuth layer with hydrogen can be separated from the bulk. After a certain time, BiH monolayer might form.

The MD simulation indicates that BiX monolayers will deform to some degree in condition partial concentration of oxygen gas is high at RT, but their honeycomb structure can remain (See details in Supplementary Information (Ⅱ)). BiX monolayers should be protected under vacuum or using inert gases environment or an anti-oxidization layer such as two-dimensional graphene, BN, or $MoS_2$. It should be pointed out that application of the strain can further modify the band gaps of the BiX/SbX monolayers. For example, a strain of 5% can increase the band gap of BiH monolayer by 0.06 eV (See Fig. S5).

In conclusion, we have identified a new family of huge-gap 2D TI phase BiX/SbX monolayers (X = H, F, Cl and Br) by FP calculations, especially BiH and BiF monolayers with known largest bulk band gaps (>1.0 eV) that far exceed the gaps of the current experimentally realized 2D TI materials. The topological characteristic of these TIs is confirmed by the calculated nontrivial $Z_2$ index and an explicit construction of the low energy effective model in the system. These giant-gaps are entirely due to the result of strong spin-orbit interaction being related to the $p_x$ and $p_y$ orbitals of the Bi/Sb atoms around the two valley $K$ and $K'$ of honeycomb lattice, which is sufficiently large for the practical application at RT. The newly discovered BiX monolayers structure survives even at a temperature of 600 K. These results represent a significant advance in TIs study, and they are expected to stimulate further work to synthesize, characterize, and utilize these new 2D TIs for fundamental exploration and practical applications at RT. Besides, the biased BiX/SbX monolayers become a quantum valley Hall insulator, and valley-selective circular dichroism is available. We find a strong coupling between the real spin and the valley pseudo-spin, which is induced by the large SOC and has modified the valley magnetic moments.

# References

1       Hasan, M. Z. & Kane, C. L. Colloquium: topological insulators. *Rev. Mod. Phys.* 82, 3045 (2010).

2       Qi, X.-L. & Zhang, S.-C. Topological insulators and superconductors. *Rev. Mod. Phys.* 83, 1057 (2011).

3       Yan, B. & Zhang, S.-C. Topological materials. *Rep. Prog. Phys.* 75, 096501 (2012).

4       Kane, C. L. & Mele, E. J. Quantum spin Hall effect in graphene. *Phys. Rev. Lett.* 95, 226801 (2005).

5       Kane, C. L. & Mele, E. J. $Z_2$ topological order and the quantum spin Hall effect. *Phys. Rev. Lett.* 95, 146802 (2005).

6       Yao, Y. G., Ye, F., Qi, X. L., Zhang, S. C. & Fang, Z. Spin-orbit gap of graphene: First-principles calculations. *Phys. Rev. B* 75 (2007).

7       Bernevig, B. A., Hughes, T. L. & Zhang, S. C. Quantum spin Hall effect and topological phase transition in HgTe quantum wells. *Science* 314, 1757-1761 (2006).

8       Konig, M. *et al.* Quantum spin hall insulator state in HgTe quantum wells. *Science* 318, 1148047 (2007).

9       Knez, I., Du, R. R. & Sullivan, G. Andreev Reflection of Helical Edge Modes in InAs/GaSb Quantum Spin Hall Insulator. *Phys. Rev. Lett.* 109, 186603 (2012).

10      Murakami, S. Quantum spin Hall effect and enhanced magnetic response by spin-orbit coupling. *Phys. Rev. Lett.* 97, 236805 (2006).

11      Liu, Z. *et al.* Stable Nontrivial $Z_2$ Topology in Ultrathin Bi(111) Films: A First-Principles Study. *Phys. Rev. Lett.* 107, 136805 (2011).

12      Weeks, C., Hu, J., Alicea, J., Franz, M. & Wu, R. Q. Engineering a Robust Quantum Spin Hall State in Graphene via Adatom Deposition (vol X1, 021001, 2011). *Physical Review X* 2, 029901 (2012).

13      Xu, Y. *et al.* Large-Gap Quantum Spin Hall Insulators in Tin Films. *Phys. Rev. Lett.* 111, 136804 (2013).

14      Wang, Z. F., Liu, Z. & Liu, F. Organic topological insulators in organometallic lattices. *Nat. Commun.* 4, 2451 (2013).

15      Liu, C. C., Feng, W. X. & Yao, Y. G. Quantum Spin Hall Effect in Silicene and Two-Dimensional Germanium. *Phys. Rev. Lett.* 107, 076802 (2011).

16      Liu, C. C., Jiang, H. & Yao, Y. G. Low-energy effective Hamiltonian involving spin-orbit coupling in silicene and two-dimensional germanium and tin. *Phys. Rev. B* 84, 195430 (2011).

17      Yang, F. *et al.* Spatial and Energy Distribution of Topological Edge States in Single Bi(111) Bilayer. *Phys. Rev. Lett.* 109, 016801 (2012).

18      Yan, B. H., Jansen, M. & Felser, C. A large-energy-gap oxide topological insulator based on the superconductor BaBiO3. *Nat. Phys.* 9, 709-711 (2013).

19      Togo, A., Oba, F. & Tanaka, I. First-principles calculations of the ferroelastic transition between rutile-type and $CaCl_2$-type $SiO_2$ at high pressures. *Phys. Rev. B* 78, 134106 (2008).

20      Gillespie, R. J. P., J. (1975). Emeleus, H. J. & Sharp, A. G. ed. Advances in inorganic chemistry & radiochemistry. Academic Press. pp.77-78. ISBN 0- 12-023617-6.

21  Greenwood, N. N. E., A. (1997). Chemistry of the elements (2nd Edn.), Oxford: Butterworth-Heinemann. ISBN 0-7506-3365-4.

22  Fu, L. & Kane, C. L. Time reversal polarization and a $Z_2$ adiabatic spin pump. *Phys. Rev. B* 74, 195312 (2006).

23  Feng, W. X., Wen, J., Zhou, J. J., Xiao, D. & Yao, Y. G. First-principles calculation of $Z_2$ topological invariants within the FP-LAPW formalism. *Comput. Phys. Commun.* 183, 1849-1859 (2012).

24  Fukui, T. & Hatsugai, Y. Quantum spin Hall effect in three dimensional materials: Lattice computation of $Z_2$ topological invariants and its application to Bi and Sb. *J. Phys. Soc. Jpn.* 76, 053702 (2007).

25  Zhang, H. J. *et al.* Topological insulators in $Bi_2Se_3$, $Bi_2Te_3$ and $Sb_2Te_3$ with a single Dirac cone on the surface. *Nat. Phys.* 5, 438-442 (2009).

26  Moore, J. Topological insulators: The next generation. *Nat. Phys.* 5, 378-380 (2009).

27  Wu, C. J., Bergman, D., Balents, L. & Sarma, S. D. Flat bands and Wigner crystallization in the Honeycomb optical lattice. *Phys. Rev. Lett.* 99 (2007).

28  Wu, C. J. Orbital Analogue of the Quantum Anomalous Hall Effect in p-Band Systems. *Phys. Rev. Lett.* 101 (2008).

29  Xiao, D., Yao, W. & Niu, Q. Valley-contrasting physics in graphene: magnetic moment and topological transport. *Phys. Rev. Lett.* 99, 236809 (2007).

30  Xiao, D., Liu, G.-B., Feng, W., Xu, X. & Yao, W. Coupled spin and valley physics in monolayers of $MoS_2$ and other group-VI dichalcogenides. *Phys. Rev. Lett.* 108, 196802 (2012).

31  Xu, X., Yao, W., Xiao, D. & Heinz, T. F. Spin and pseudospins in layered transition metal dichalcogenides. *Nat. Phys.* 10, 343-350, (2014).

32  Godfrey, S. M., McAuliffe, C. A., Mackie, A.G. & Pritchard, R. G. Nicholas, C. N. ed. (1998), Chemistry of arsenic, antimony, & bismuth. Springer. pp. 67-84. ISBN 0-7514-0389-X.

33  Wiberg, E., Wiberg, N. & Holleman, A. F. (2001), Inorganic chemistry. Academic Press. ISBN 0-12-352651-5.

34  Elias, D. *et al.* Control of graphene's properties by reversible hydrogenation: evidence for graphane. *Science* 323, 610-613 (2009).

35  Balog, R. *et al.* Bandgap opening in graphene induced by patterned hydrogen adsorption. *Nat. Mater.* 9, 315-319 (2010).

36  Sun, Z. *et al.* Towards hybrid superlattices in graphene. *Nat. Commun.* 2, 559 (2011).

37  Ryu, S. *et al.* Reversible basal plane hydrogenation of graphene. *Nano Lett.* 8, 4597-4602 (2008).

38  Pakdel, A., Bando, Y. & Golberg, D. Nano boron nitride flatland. *Chem. Soc. Rev.* 43, 934-959 (2014).

**Acknowledgments**

This work was supported by the MOST Project of China (Nos. 2014CB920903, 2010CB833104, 2011CBA00100, and 2013CB932604 ), the National Natural Science Foundation of China (Nos. 11174337, 11225418, 51371009, 50971003, 51171001, and 11274016), the National High Technology Research and Development Program of China (No. 2011AA03A403), the Specialized Research Fund for the Doctoral Program of Higher Education of China (No. 20121101110046, 20130001110002), and Excellent young scholars Research Fund of Beijing Institute of Technology (No. 2013YR1816).

**Authors contributions:** † These authors contributed equally to this work. Z.G.S., J.B.Y., J.L., & Y.G.Y. conceived the research. Z.G.S., B.T.F. & M.Y. performed the calculations. C.C.L., & Z.G.S. performed the analysis. Z.G.S., C.C.L., J.B.Y., J.L. and Y.G.Y. wrote the manuscript. J.Z.H, Y.C.Y, Q.N., J.L & Y.G.Y. contributed in the discussions and editing of the manuscript.

**Additional Information**

Supplementary Information accompanies this paper at http://www.natureasia.com

**Competing financial interests:**

The authors declare no competing financial interests.

**Corresponding authors:**

Correspondence and requests for materials should be addressed to J.Y. (email: jbyang@pku.edu.cn) or to J.L. (email: jinglu@pku.edu.cn) or to Y.Y. (email: ygyao@bit.edu.cn).

Table 1: The lattice constant $a$, Bi-Bi and Sb-Sb bond length $b$, Bi-X and Sb-X bond length $d$, buckling height $h$ shown in Fig. 1(a) ($h$ defined as the distance from the center of upper to that of lower Bi/Sb atoms), global band gap $\Delta$ (Superscripts $d$ and $i$ represent the direct gap and the indirect gap, respectively). Fermi velocity $v_F$ and the magnitude of the intrinsic effective SOC $\lambda_{so}$ for Bi and Sb hydride/halide monolayers, which are obtained from the FP-calculations. Note that $\lambda_{so} = E_g/2$, with local gap $E_g$ opened by SOC at the Dirac point.

| Monolayers | $a$ (Å) | $b$ (Å) | $d$ (Å) | $h$ (Å) | $\Delta$ (eV) | $v_F(10^5 m/s)$ | $\lambda_{so}$ | $Z_2$ |
|---|---|---|---|---|---|---|---|---|
| BiH | 5.53 | 3.19 | 1.82 | 0.08 | $1.03^i$ | 8.9 | 0.56 | 1 |
| BiF | 5.38 | 3.14 | 2.12 | 0.46 | $1.08^i$ | 7.2 | 0.55 | 1 |
| BiCl | 5.49 | 3.17 | 2.54 | 0.24 | $0.95^i$ | 6.9 | 0.56 | 1 |
| BiBr | 5.52 | 3.18 | 2.69 | 0.16 | $0.74^i$ | 8.0 | 0.65 | 1 |
| SbH | 5.29 | 3.05 | 1.73 | 0.08 | $0.41^d$ | 8.6 | 0.21 | 1 |
| SbF | 5.12 | 2.96 | 1.99 | 0.30 | $0.32^d$ | 7.9 | 0.16 | 1 |
| SbCl | 5.17 | 2.98 | 2.42 | 0.14 | $0.36^d$ | 7.7 | 0.18 | 1 |
| SbBr | 5.25 | 3.03 | 2.57 | 0.09 | $0.40^d$ | 8.6 | 0.20 | 1 |

Table 2: Comparisons between Gibbs free energies of BiX/SbX monolayers and those of BiX$_3$/ SbX$_3$ molecules.

| Monolayer | $\delta G$ (eV) | Molecule | $\delta G$ (eV) |
|:---:|:---:|:---:|:---:|
| BiH | 0.299 | BiH$_3$ | 0.53 |
| BiF | -1.23 | BiF$_3$ | -1.976 |
| BiCl | -0.509 | BiCl$_3$ | -0.792 |
| BiBr | -0.412 | BiBr$_3$ | -0.899 |
| SbH | 0.219 | SbH$_3$ | 0.264 |
| SbF | -1.288 | SbF$_3$ | -2.204 |
| SbCl | -0.478 | SbCl$_3$ | -0.858 |
| SbBr | -0.374 | SbBr$_3$ | -0.579 |

**Figure captions**

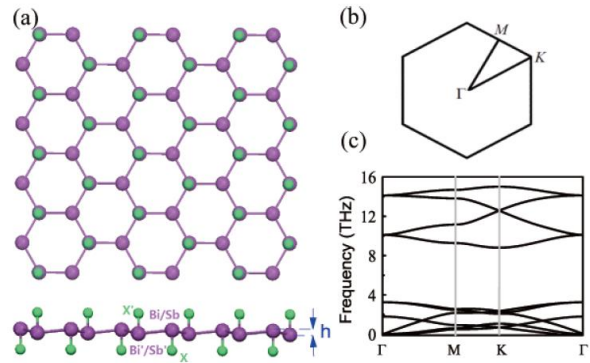

Figure 1: (a) Lattice geometry for BiX/SbX monolayer (X = H, F, Cl, and Br) from the top view (upper) and side view (lower), respectively. In a unit cell BiX/SbX is related to Bi'X'/Sb'X' by an inversion operation. (b) First Brillouin zone of BiX/SbX monolayers and the points of high symmetry. (c) Corresponding phonon spectrum for BiH monolayer.

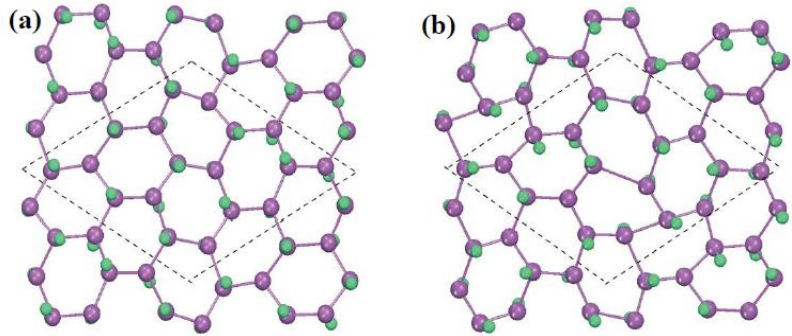

Figure 2: Snapshots from the MD simulation of the structure for BiH monolayer at the temperatures of 300 K (a) and 600 K (b) after 2.25 ps. Pink balls: Bi atoms; green balls: H atoms, and the dashed line indicates a supercell with $3 \times 3$ unit cell.

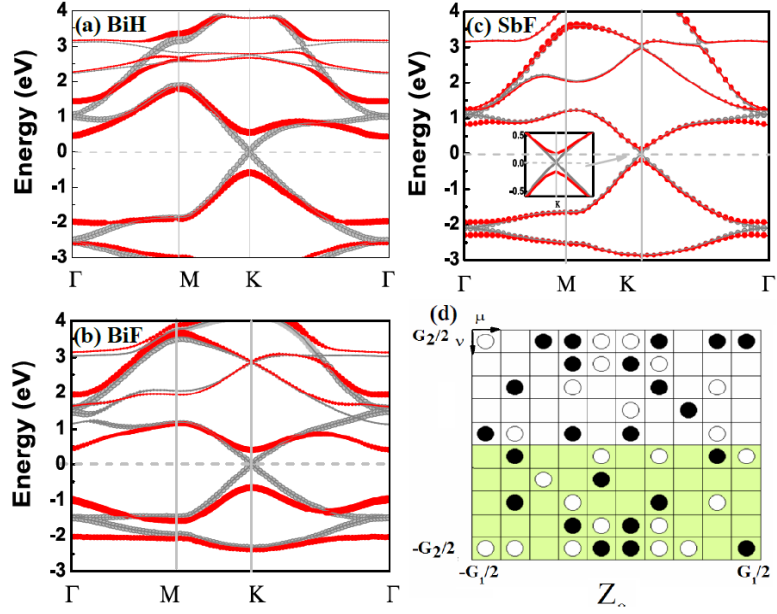

Figure 3: Band structures of the BiX monolayers without (gray) and with (red) including of the SOC and $Z_2$ topological invariant: (a) BiH, (b) BiF, and (c) SbF monolayers. The Fermi level is set to zero. The bands near the Fermi level consist of the $p_x$ and $p_y$ orbitals. The size of the symbols is proportional to the population of the $p_x$ and $p_y$ orbitals. (d) $n$-field configuration for BiH monolayer. The calculated torus in Brillouin zone is spanned by $G_1$ and $G_2$. Note that the two reciprocal lattice vectors form an angle of 120 degrees. The white and black circles denote $n = 1$ and $-1$, respectively, while the blank denotes 0. The $Z_2$ invariant is 1 obtained by summing the $n$-field over half of the torus mod 2.

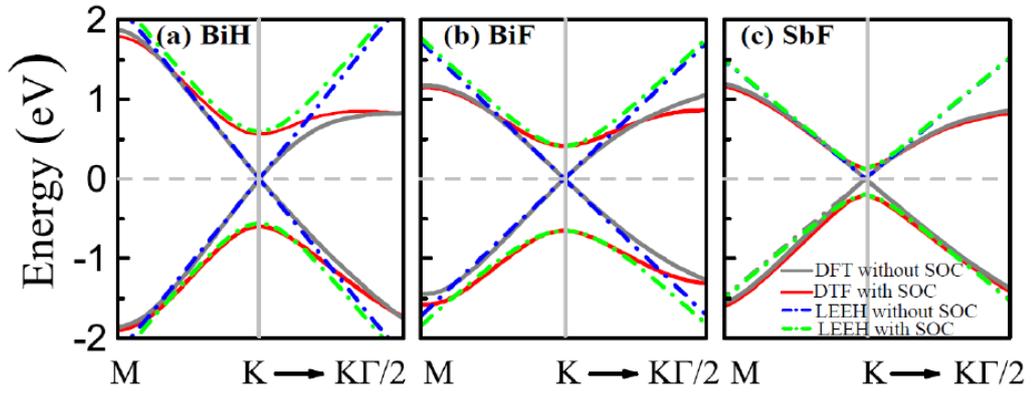

Figure 4: A comparison of the band structures round the *K* point for (a) BiH, (b) BiF, and (c) SbF monolayers calculated using DFT and LEEH methods. Solid gray and solid red lines denote the data calculated from DFT theory without SOC and with SOC, respectively. Dashed blue and dashed green lines represent the data calculated from LEEH method with SOC and without SOC, respectively. The Fermi level is set to be zero.

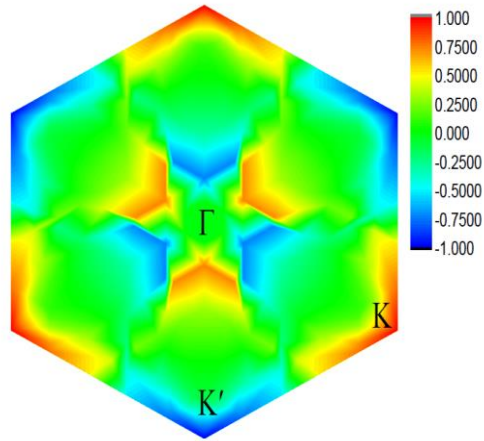

Figure 5: Degree of circular polarization $\eta(\mathbf{k})$ in irreducible Brillouin zone of BiH monolayer under a vertical electric field.